\documentclass[aps,prb,superscriptaddress,footinbib,longbibliography]{revtex4-1}
\usepackage{graphicx}
\usepackage{amsmath}
\usepackage{color}
\usepackage[normalem]{ulem}
\usepackage[colorlinks, linkcolor= blue, citecolor = blue, urlcolor=blue]{hyperref}

\begin{document}

\title{Nonlocal effects in temporal metamaterials}
\author{Carlo Rizza} \email{carlo.rizza@univaq.it}
\affiliation{Department of Physical and Chemical Sciences, University of L'Aquila, Via Vetoio 1, I-67100 L'Aquila, Italy}
\author{Giuseppe Castaldi}
\affiliation{University of Sannio, Department of Engineering, Fields \& Waves Lab, Benevento, I-82100, Italy} 
\author{Vincenzo Galdi} \email{vgaldi@unisannio.it}
\affiliation{University of Sannio, Department of Engineering, Fields \& Waves Lab, Benevento, I-82100, Italy}

	
\begin{abstract}
Nonlocality is a fundamental concept in photonics. For instance, nonlocal wave-matter interactions in spatially modulated metamaterials enable novel effects, such as giant electromagnetic chirality, artificial magnetism, and negative refraction. Here, we investigate the effects induced by spatial nonlocality in {\em temporal} metamaterials, i.e., media with a dielectric permittivity rapidly modulated in time. Via a rigorous multiscale approach, we introduce a general and compact formalism for the nonlocal effective medium theory of temporally periodic metamaterials. In particular, we study two scenarios: {\em i)} a periodic temporal modulation, and {{\em ii)}} a temporal boundary where the permittivity is abruptly changed in time and subject to periodic modulation. We show that these configurations can give rise to peculiar nonlocal effects, and we highlight the similarities and differences with respect to the spatial-metamaterial counterparts. Interestingly, by tailoring the effective boundary wave-matter interactions, we also identify an intriguing configuration for which a temporal metamaterial can perform the first-order derivative of an incident wavepacket. Our theoretical results, backed by full-wave numerical simulations, introduce key physical ingredients that may pave the way for novel applications. By fully exploiting the time-reversal symmetry breaking, nonlocal temporal metamaterials promise a great potential for efficient, tunable optical computing devices.
\end{abstract}

\maketitle

\section{Introduction} 

Spatial dispersion \cite{Landau,Agranovich} implies that the electromagnetic (EM) constitutive relationships are {\em nonlocal}, i.e.,
the electric and/or magnetic inductions at a given point also depend on the fields applied in a spatial neighborhood (and, because of causality, at previous time instants). From a mathematical viewpoint, this can be modeled in terms of spatial derivatives in the constitutive relationships or, equivalently, via wavevector-dependent constitutive parameters in the reciprocal domain.
Such behavior is typically negligible in most natural materials; nevertheless it can become a dominant effect in artificial materials, such as metamaterials and photonic crystals, characterized by {\em spatially periodic} (or almost periodic) arrangements of basic constituents \cite{Elser,Silveirinha2,Chebykin,Rizza_PRB}. Although in some cases nonlocality is viewed as a detrimental effect to suppress or mitigate \cite{Pendry}, its proper harnessing can be very beneficial in a variety of application scenarios, including artificial magnetism \cite {Alu}, chirality \cite{Rizza_PRB}, 
ultrafast nonlinear optics \cite{Wurtz},
advanced dispersion engineering \cite{Castaldi,Moccia}, 
and wave-based analog computing \cite{Silva}.

Currently, in metamaterials engineering, there is a surge of interest in exploiting the {\em temporal} dimension as well, motivated by the increasing availability of fast, reconfigurable ``meta-atoms'' whose response can be dynamically modulated in time \cite{Shaltout,Caloz1,Caloz2,Engheta}. This has led to revisiting with renewed attention some old studies on wave interactions with time-varying media or structures \cite{Morgenthaler,Oliner,Fante}, and to the demonstration of a variety of intriguing effects and applications, ranging from nonreciprocity to broadband light manipulation (see, e.g., \cite{Hadad, Shaltout1, Bacot, Shlivinski, Huidobro, Galiffi1,Mekawy, Ramaccia, Galiffi, Barati, Engheta_1,Torrent,Engheta_2, Engheta_3,Mosallaeia2,Li, Ramaccia1, Castaldi1,Galiffi2,Mosallaeia1} for a sparse sampling).

Interestingly, by exploiting the space-time duality, the concept of effective medium theory (EMT) has been translated from conventional spatial multilayers to temporal ``multisteps'' featuring a time-varying permittivity \cite{Engheta_1}, and higher-order homogenization schemes have also been put forward to study nonlocal effects \cite{Torrent}.

In this paper, we revisit these concepts via first-principle calculations based on a multiscale approach  \cite{Rizza_PRB}. We show that nonlocality in temporal metamaterials can induce an effective diamagnetic response, in analogy with the nonlocal effects observed in conventional spatial metamaterials of infinite extent. Moreover, in analogy with the boundary-type nonlocal effects observed in truncated spatial metamaterials, we also consider a temporal scenario  
where the permittivity of an unbounded nondispersive medium is abruptly changed in time and subject to a temporally periodic modulation. Remarkably, we show that this temporal boundary can give rise to peculiar nonlocal effects which, in suitably tailored parameter regimes, can be harnessed to perform elementary analog-computing operations, such as computing the first-order derivative of an incident wavepacket. Finally, for validation, we also carry out full-wave numerical simulations, which are in good agreement with our theoretical derivations. Our results indicate that nonlocality in temporal metamaterials may play a key role in engineering novel effects in nanophotonics and optical computing.

\section{Results}

\subsection{Nonlocal effective medium theory}
\label{EMT}
Let us consider an isotropic, generally inhomogeneous, time-modulated medium, whose EM response is described by a relative dielectric permittivity periodically modulated in time, $\varepsilon({\bf r},t)=\varepsilon({\bf r},t+\tau)$. 
In our study, we assume that the operating frequencies are much lower than any material resonance frequencies, so that temporal dispersion effects can be approximately neglected \cite{Morgenthaler,Solis}.
From Maxwell's equations, the field dynamics can be described by the vector wave equation for the electric induction $\bf D$, namely
\begin{eqnarray}
\label{eq_D}
\frac{\partial^2 {\bf D}}{\partial t^2} + c^2 {\hat L} \left[ \varepsilon^{-1}({\bf r},t) {\bf D} \right]=0,
\end{eqnarray}
where ${\hat L}=\nabla \times \nabla \times$ is the double-curl operator, and $c$ denotes the wavespeed in vacuum \cite{Torrent}. 
In the presence of a rapid temporal modulation (i.e., with a modulation angular frequency $\Omega=2 \pi/\tau$ much higher than the EM carrier one $\omega$), it is convenient to introduce the parameter $\eta=\omega/\Omega$. By exploiting the standard multiscale approach \cite{Rizza_PRB}, we can develop an asymptotic
analysis in the regime where $\eta \ll 1$. It is natural to assume for the EM observables a separate dependence on the {\em slow} and {\em fast} temporal scales ($t$ and $T=t/\eta$, respectively), and to represent them as a Maclaurin series expansion in the scale-parameter $\eta$, i.e., 
\begin{eqnarray}
A({\bf r},t,T) =\overline{A}({\bf r},t)+ \widetilde{A} ({\bf r},t,T),
\end{eqnarray}
with $\overline{A}=\sum_{m=0}^{+\infty} \eta^{m}\overline{A}^{(m)}$ and $\widetilde{A}=\sum_{m=0}^{+\infty} \eta^{m} \widetilde{A}^{(m)}$. Here, $A$ represents a generic field component, the superscript $(m)$ labels the order of each term, whereas the overline and the tilde label the {\em averaged} and {\em  rapidly varying} contributions to each order, respectively. By noticing that the relative dielectric permittivity only depends on the fast coordinate $T$, and by substituting the multiscale series expansion in Equation (\ref{eq_D}) (up to the second order in $\eta$), we obtain   
\begin{eqnarray}
	\label{eff1}
&& \frac{\partial^2 {\bf \overline{D}}}{\partial t^2}+ c^2 {\hat L} \left[ \varepsilon_{eff}^{-1} {\bf \overline{D}} + c^2 \eta^2 \sum_{n \neq 0} \frac{a_{-n}}{n^2 {\widetilde \Omega}^2 } {\hat L} \left( a_n {\bf \overline{D}} \right)\right]=0, \nonumber \\
&& \widetilde{\bf{D}}=c^2   \sum_{n \neq 0}  \frac{e^{ i n {\widetilde \Omega} T }}{n^2 {\widetilde \Omega}^2} {\hat L}  \left( a_n \overline{\bf D} \right),
\end{eqnarray}
where we have separated the fast and slow contributions, and we have assumed that the reciprocal of the relative permittivity admits the Fourier series expansion
\begin{equation}
\label{Fourier} 
\varepsilon^{-1}({\bf r},T)=\sum_{n=-\infty}^{+\infty} a_n ({\bf r}) e^{i n {\widetilde \Omega} T },
\end{equation}
with $\varepsilon_{eff}^{-1}=a_0$, ${\widetilde \Omega}= \eta \Omega$, and $i$ denoting the imaginary unit. 
Equivalently, we can write Maxwell's equations governing the dynamics of the average fields in our temporal metamaterial as 
\begin{eqnarray}
\label{Max_eff} 
\nabla \cdot \overline{\bf D}&=&0, \quad \quad 
\nabla \cdot \overline{\bf B}=0, \nonumber \\
\nabla \times \overline{\bf E} &=& -\frac{\partial {\bf \overline{ \bf B}}}{\partial t}, \quad \quad 
\nabla \times \overline{\bf H} = \frac{\partial {\bf \overline{ \bf D}}}{\partial t}, \nonumber \\
\end{eqnarray}
along with the constitutive relationships
\begin{eqnarray}
\label{cost} 
\overline{\bf E} &=& \varepsilon_0^{-1} \left[ \varepsilon_{eff}^{-1} {\bf \overline{D}} + c^2 \eta^2 \sum_{n \neq 0} \frac{a_{-n}}{n^2 {\widetilde \Omega}^2 } {\hat L} \left( a_n {\bf \overline{D}} \right)\right], \nonumber \\
\overline{\bf B} &=& \mu_0 \overline{\bf H}, 
\end{eqnarray}
which are evidently {\em nonlocal} because of the presence of field derivatives.
By considering the limit $\eta \rightarrow 0$, it is apparent that the term proportional to $\eta^2$ in the first of Equations (\ref{eff1}) vanishes, and the temporal metamaterial behaves as a dielectric medium with an effective relative permittivity $\varepsilon_{eff}$, thereby recovering the results in previous studies \cite{Engheta_1,Torrent}. 
In the simplest case of spatial homogeneity (i.e., spatially unbounded, time-varying media \cite{Engheta_1}), the first of Equations (\ref{cost}) becomes 
\begin{eqnarray}
\label{cost2} 
\overline{\bf E} &=& \varepsilon_0^{-1} \left( \varepsilon_{eff}^{-1} {\bf \overline{D}} - \frac{\gamma}{K^2}  \nabla^2 {\bf \overline{\bf D}}\right),  
\end{eqnarray}
where $K=2 \pi /(c \tau)$ and $\gamma=2 \sum_{n =1}^{+\infty} |a_n|^2 /n^2$.

Next, we consider a monochromatic plane wave propagating in a temporal metamaterial, i.e., $\overline{\bf D}=2 {\rm Re} \left[ {\bf d}_0  e^{i kz -i \omega(k) t} \right]$. By substituting this expression in the effective Maxwell's equations, and exploiting the constitutive relationship in Equation (\ref{cost2}), we derive the dispersion relationship 
\begin{equation}
\omega^2(k)=c^2 \frac{k^2}{\varepsilon_{eff} \mu_{eff}(k)},
\end{equation}
where 
\begin{equation}
\label{muu}
\mu_{eff}(k)=\frac{1}{1+\gamma \varepsilon_{eff} \frac{k^2}{K^2}}
\end{equation}  
plays the role of an effective relative magnetic permeability.
Note that artificial magnetism in time-modulated dielectric media has been predicted in a recent study \cite{Torrent}, suggesting that a temporal metamaterial can behave as a resonant magneto-dielectric medium. Our results above  confirm the previous findings, and establish {\em tighter} bounds on the attainable magnetic response. Specifically, for a temporally periodic metamaterial based on positive-permittivity modulation [$\varepsilon(t)=\varepsilon(t+\tau)>0$], it is apparent from Equation (\ref{muu}) (recalling that $\gamma\ge 0$) that only nonlocal {\em diamagnetism} with $0<\mu_{eff}(k)<1$ can be attained, whereas resonant, paramagnetic, and $\mu$-negative responses are {\em forbidden}. 

We highlight that, as in conventional spatially modulated metamaterials, optical magnetism stems from spatial dispersion. As previously mentioned, in spatial metamaterials, nonlocality can strongly affect the EM response, and it can produce undesirable effects \cite{Pendry}. Conversely, in an isotropic temporal metamaterial, the spatial dispersion up to the second order (i.e., up to $\eta^2$) is {\em fully equivalent} to optical magnetism described by the effective relative magnetic permeability $\mu_{eff}(k)$ in Equation (\ref{muu}) (see the Methods Section \ref{TOM} for further details). 
\begin{figure*}
\centering
\includegraphics[width=1\textwidth]{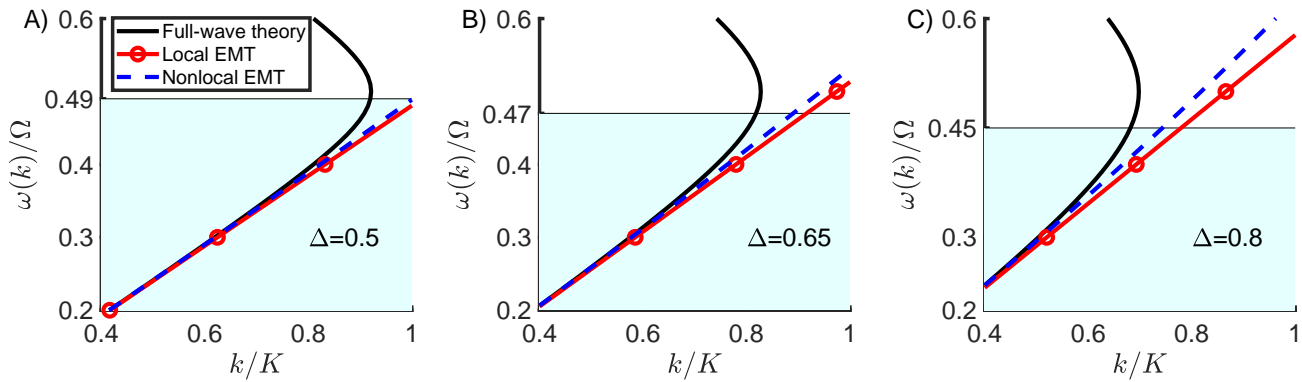}
\caption{Comparison between the predictions from the full-wave theory and both local an nonlocal  EMT models for a temporal metamaterial with relative permittivity as in Equation (\ref{e(t)}), with $\overline{\varepsilon}=5$. 
	(A), (B), (C) Normalized angular frequency $\omega/\Omega$ as function of the normalized wavenumber $k/K$, for $\Delta=0.5, 0.65$, and $0.8$, respectively.The light-blue shaded area indicates the region where the nonlocal EMT works well ($\lesssim 10\%$ error with respect to full-wave theory). 
\label{comp1}}
\end{figure*}

For some quantitative assessments, we assume that the relative permittivity profile is given by 
\begin{equation}
\label{e(t)}
\varepsilon(t)=\overline{\varepsilon} \left[1 +\Delta \cos(2 \pi t/\tau+ \phi) \right],
\end{equation}
with $\Delta$ parameterizing the modulation depth, and $\phi$ being an arbitrary phase.
To validate and calibrate the predictions of our proposed nonlocal EMT model, we compare them with the results from a rigorous full-wave theory (see the Methods Section \ref{TOM} for further details), as well as with the conventional (local) EMT \cite{Engheta}. 
Figure \ref{comp1} compares these three predictions for the dispersion relationship $\omega(k)$, for $\overline{\varepsilon}=5$, and three different values of the modulation depth $\Delta$.  Here and henceforth, consistently with our assumption of a negligible temporal dispersion, parameters are chosen so that the relative permittivity in time is always greater than one.
Note that the effective parameters $\varepsilon_{eff}$ and $\mu_{eff}(k)$ do not depend on $\phi$, since this latter merely corresponds to a time shift. As expected, our nonlocal EMT is in fair agreement with the full-wave results and the conventional (local) EMT in the quasi-homogenized (weak-dispersion) regime $k/K \ll 1$, and we observe that it works well within the regions $\omega(k)/\Omega<0.45,0.47,0.49$ for $\Delta=0.5,0.65,0.8$, respectively, as shown in Figure \ref{comp1}.
Note that $\omega(k)/\Omega$ corresponds to the scale parameter $\eta$ that rules the multiscale homogenization process (see Section \ref{EMT}). By increasing the modulation depth $\Delta$, we expect the nonlocal effects to become more pronounced. This is evident in the three panels of Figure \ref{comp1}, as the wavenumber region where the temporal metamaterial exhibits a homogeneous behavior [i.e., where $\omega(k)$ is approximately linear] progressively shrinks as $\Delta$ increases. 

\begin{figure}
\centering
\includegraphics[width=0.5\textwidth]{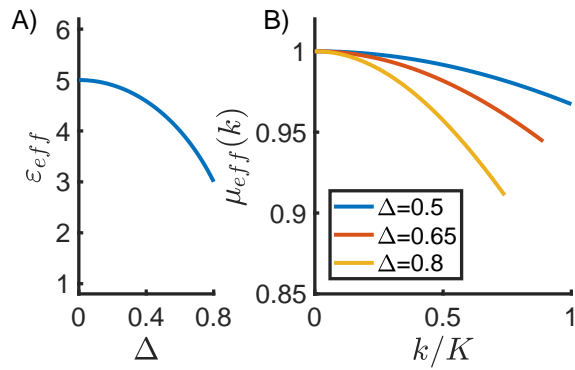}
\caption{Effective parameters for the temporal metamaterial considered in Figure \ref{comp1}. (A) Effective relative  permittivity $\epsilon_{eff}$ as a function of $\Delta$;   (B) Effective relative  permeability $\mu_{eff}(k)$ as a function of $k/K$, for different values of the modulation depth $\Delta$. Note that $\mu_{eff}(k)$ is only shown within the region where the nonlocal EMT holds ($\lesssim 10\%$ error with respect to full-wave theory).
\label{fig_homo}}
\end{figure}

Figure \ref{fig_homo}(A) shows the effective relative permittivity $\varepsilon_{eff}$ as a function of $\Delta$, whereas Figure \ref{fig_homo}(B) shows the effective relative permeability $\mu_{eff}(k)$ as a function of $k/K$, for different values of $\Delta$; these effective parameters are only shown within the region where the nonlocal EMT is in good agreement with the full-wave theory (corresponding to the light-blue shaded area in Figure \ref{comp1}). We observe that, in the case of a deep temporal modulation (e.g., $\Delta = 0.8$), the metamaterial exhibits a significant diamagnetic response (e.g., $\mu_{eff}\simeq 0.91$ for $k/K=0.73$).   

\subsection{Nonlocal temporal boundary}
In analogy with the spatial scenario, the results derived in Section \ref{EMT} play the role of the ``bulk'' response of an infinite-extent metamaterial. Since a series  of recent studies  on {\em spatially truncated} multilayered dielectric metamaterials (see, e.g., \cite{Segev,Castaldi3,Gorlach}) have  indicated the possible onset of intriguing ``boundary'' effects in critical parameter regimes, it appears suggestive to explore similar effects in our temporal scenario here. To this aim,  we derive the conditions at the boundary of a nonlocal temporal metamaterial. More precisely, we consider a spatially homogeneous, unbounded, temporal metamaterial exhibiting a temporal boundary at a given time instant $t=t_0$, at which the relative permittivity abruptly changes from a constant value $\varepsilon_1$ to a time-varying periodic function
$\varepsilon(t)$, viz., 
\begin{eqnarray}
\label{temp}
\varepsilon_b(t)=
\left \{ \begin{array}{ll}
 \varepsilon_1, \quad t< t_0, \\
  \varepsilon(t), \quad t > t_0.
\end{array}
\right.
\end{eqnarray}
Obviously, since the medium response to modulation cannot be  infinitely fast, the assumption of abrupt temporal transitions is highly idealized, and a finite rise/fall time should be considered. However, for fall/rise times much shorter than the period of the incident wave, our approach remains valid. Within this framework, it is also worth pointing out that, in our numerical simulations (see the Methods Section \ref{fullwave}), we actually assume {\em smooth} transitions with sufficiently short rise/fall times in order to favor convergence.
As for the canonical temporal boundary \cite{Xiao}, where the dielectric permittivity exhibits a temporal transition between two constant values, the \textit{microscopic} electric induction $\bf D$ and magnetic induction $\bf B$ remain continuous across the boundary. Here, we consider plane waves, ${\bf D}=2 {\rm Re}[ {\bf d}({\bf k}, t) e^{i {\bf k} \cdot {\bf r}}]$ and ${\bf B}=2 {\rm Re}[ {\bf b} ({\bf k},t) e^{i {\bf k} \cdot {\bf r}}]$, experiencing the temporal boundary described by Equation (\ref{temp}). By enforcing the standard boundary conditions \cite{Xiao}, and taking into account the second of Equations (\ref{eff1}) and the microscopic Maxwell's curl equation $\nabla \times {\bf B} =\mu_0 \frac{\partial {\bf D}}{\partial t}$, we obtain 
\begin{eqnarray}
\label{bound}
{\bf d}_- &=& \left(1+\alpha_0 \frac{k^2}{K^2}   \right) {\bf d}_+, \nonumber \\
{\bf b}_- &=& \left(1+\alpha_0 \frac{k^2}{K^2}   \right) {\bf b}_+ -i c \mu_0 \beta_0  \frac{\bf k}{K} \times {\bf d}_+.  \nonumber  \\
\end{eqnarray}
where ${\bf d}_{\pm}=\overline{{\bf d}}({\bf k},t_0^{\pm})$ and ${\bf b}_{\pm}=\overline{{\bf b}}({\bf k},t_0^{\pm})$, with 
\begin{eqnarray}
\label{ab}
\alpha_0&=&2 {\rm Re} \left( \sum_{n =1}^{+\infty} \frac{a_n}{n^2} e^{i n \Omega t_0} \right), \nonumber \\ 
\beta_0&=&2 {\rm Im}  \left( \sum_{n =1}^{+\infty} \frac{a_n}{n}e^{i n \Omega t_0}  \right).
\end{eqnarray}
From Equations (\ref{bound}), it is evident that, in the limit $k/K \rightarrow 0$, the inductions  $\bf d$ and $\bf b$ are continuous, and our approach reproduces the known results for the canonical temporal boundary \cite{Xiao}. 
Generally, the homogenization process can hide the asymmetry of the system. In the spatial domain, the EM chirality is washed out in the conventional homogenized response of a composite metamaterial \cite{Gorlach,Rizza_PRB}. Here, the system exhibits a time-reversal symmetry breaking due to the permittivity modulation in time and, while this effect is lost in the effective ``bulk'' response,  it is restored in the boundary conditions expressed by Equations (\ref{bound}). This  is somehow analogous to what is observed in  spatial multilayered metamaterials, which exhibit chiral boundary effects attributable to the parity symmetry breaking \cite{Gorlach}.  The abrupt switching of the  permittivity  breaks the time-reversal symmetry. On the other hand, by comparison with the conventional boundary conditions (i.e., continuity of $\bf d$ and $\bf b$ at the temporal boundary), we highlight that, in Equations (\ref{bound}),  the term proportional to $\beta_0$ breaks explicitly the time-reversal symmetry. The nonlocality preserves the time-reversal asymmetry of the ``microscopic'' permittivity modulation, and this peculiar wave-matter interaction endows us with novel degrees of freedom for manipulating the wave propagation. 
\begin{figure}
\centering
\includegraphics[width=0.5\textwidth]{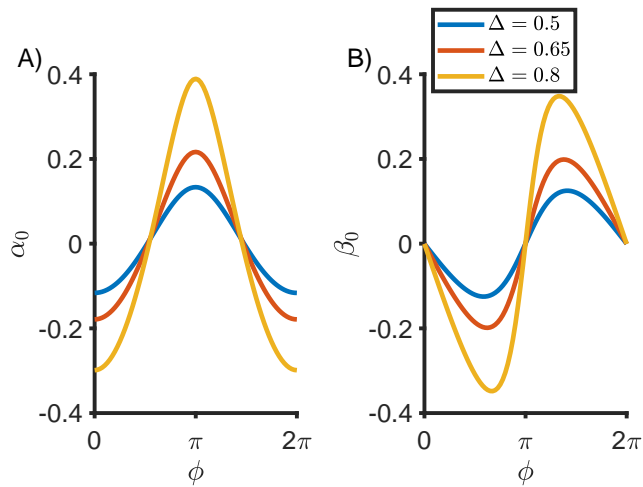}
\caption{(A), (B) Nonlocal effective boundary parameters $\alpha_0$ and $\beta_0$, respectively, as a function of  the modulation  phase $\phi$. Here, $t_0=0$, and a metamaterial with the permittivity profile in Equation (\ref{e(t)}) is considered with $\overline{\varepsilon}=5$.} 
\label{fig_alp0_bet0}
\end{figure}

\begin{figure}
\centering
\includegraphics[width=0.5\textwidth]{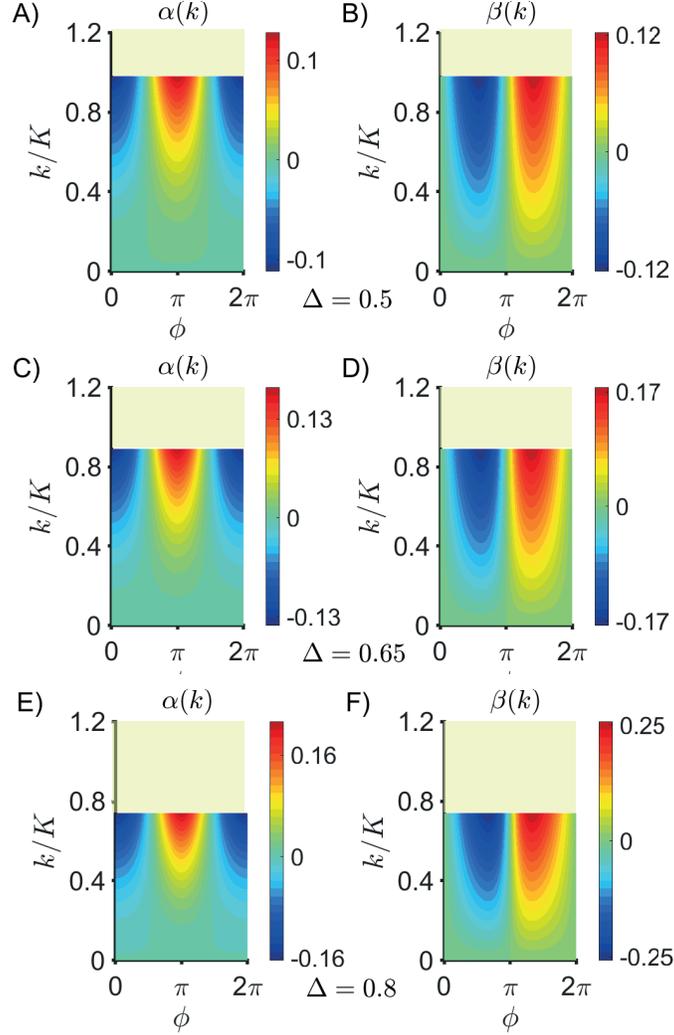}
\caption{Nonlocal contributions in Equations (\ref{nonlocal}) as a function of $k/K$ and $\phi$, for $\Delta=0.5$ (A,B), $\Delta=0.65$ (C,D), and $\Delta=0.8$ (E,F). Here, $t_0=0$ and a temporal metamaterial with the  permittivity profile in Equation (\ref{e(t)}) is considered, with $\overline{\varepsilon}=5$. The parameters are only shown within the region where the nonlocal EMT holds ($\lesssim 10\%$ error with respect to full-wave theory).} 
\label{fig_alp_bet}
\end{figure}

As an example, we consider once again the temporal metamaterial described by Equation (\ref{e(t)}), with $\overline{\varepsilon}=5$. Figure (\ref{fig_alp0_bet0}) shows the nonlocal effective parameters $\alpha_0$ and $\beta_0$ as a function of the  phase parameter $\phi$, for $\Delta=0.5,0.65,0.8$. It is evident that the EM boundary response strongly depends on  $\phi$, which thus constitutes an additional ``knob'' for tailoring the wave propagation. As a general result, it is worth noting that $\alpha_0=0$ for $\phi=\pi/2,3\pi/2$, whereas $\beta_0=0$ for $\phi=0,\pi$. By exploiting Equation (\ref{ab}), one can prove that the parameters $\alpha_0$ and $\beta_0$ vanish when the dielectric function exhibits even [$\varepsilon (t-t_0)=\varepsilon (t_0-t)$] or odd [$\varepsilon (t-t_0)=-\varepsilon (t_0-t)$] parity, respectively. Similar and related effects have been investigated in multilayered dielectric metamaterials \cite{Gorlach}, and topological photonic crystals \cite{HHH}. To better understand the impact of the novel nonlocal terms in the boundary conditions expressed by Equations (\ref{bound}), Figure (\ref{fig_alp_bet}) shows the two nonlocal contributions 
\begin{equation}
\alpha(k)=\alpha_0 k^2/K^2, \quad\beta(k)=\beta_0 k/K,
\label{nonlocal}
\end{equation}
 as a function of $k/K$ and $\phi$, for $\Delta=0.5$ [panels (A,B)], $\Delta=0.65$ [panels (C,D)], and $\Delta=0.8$ [panels (E,F)]; as in previous examples, the parameters are only shown within the region where the nonlocal EMT holds (i.e., the deviation is less than $10\%$ with respect to full-wave theory).
 Recalling the results in Figure \ref{comp1}, the larger the modulation depth $\Delta$, the smaller the wavenumber region where the nonlocal EMT is in agreement with the full-wave theory. The results in Figure (\ref{fig_alp_bet}) indicate that the impact of nonlocality at the temporal boundary increases for deeper modulations. In particular, for $\Delta=0.8$, $\alpha(k)$ and $\beta(k)$ reach the values $0.21$ and $0.26$, respectively, for $k/K \simeq 0.74$. 

\subsection{Reflection and transmission at a nonlocal temporal boundary}
Let us consider a monochromatic plane wave propagating  in a spatially homogeneous unbounded medium with a relative permittivity as described by Equation (\ref{temp}). Without loss of generality, we assume propagation along the positive $z$ direction, and an $x$-polarized  electric induction, i.e. ${\bf D}(t,z)=2 \mbox{Re}[d(k,t) e^{i k z}] \hat{\bf e}_{x}$. Following the nonlocal EMT model developed in Section \ref{EMT}, the average EM fields are governed by Equations (\ref{Max_eff}), where the effective wave-matter coupling is given by 
\begin{equation}
\label{ddd}
\overline{\bf D}=\varepsilon_0 \varepsilon_1 \overline{\bf E},
\end{equation}
for $t<t_0$,  and by Equation (\ref{cost2}) for $t>t_0$.    
As a consequence, by neglecting the fast components, the electric induction $d(k,t)$ is merely given by 
\begin{eqnarray}
\label{epp}
d(k,t)=
\left\{ \begin{array}{ll}
  d_{in}(k) e^{- i \omega_1(k) t},  \quad t< t_0, \\
  d_t(k) e^{- i \omega(k) t}+ d_r(k) e^{i \omega(k) t},  \quad t > t_0, 
\end{array}
\right.
\end{eqnarray}
where $\omega_1(k)=c k/\sqrt{\varepsilon_1}$, $\omega(k)=c k/\sqrt{\varepsilon_{eff}\mu_{eff}(k)}$, and $d_{in}$, $d_r$, $d_t$ are the incident, reflected (backward) and transmitted (forward) amplitudes, respectively. 
By means of the third of Equations (\ref{Max_eff}) and Equation (\ref{cost2}), we can derive  the magnetic field associated to the electric induction in Equation (\ref{epp}). Then, by enforcing the temporal  boundary conditions in Equations (\ref{bound}), after some straightforward algebra, we obtain
the temporal transmission [$t_d=\frac{{d}_{t}}{{d}_{in}} e^{-i [\omega(k)-\omega_1(k)] t_0}$] and reflection [$r_d=\frac{{d}_{r}}{{d}_{in}}  e^{i [\omega(k)+\omega_1(k)] t_0}$] coefficients for the electric induction

\begin{eqnarray}
\label{tr}
t_d (k)&=&\frac{n_1+n_{eff}(k)}{2 n_1 \left[1+\alpha(k)\right]}+
i \beta_0  \frac{  n_{eff}(k) }{2  \left[1+\alpha(k)\right]^2} \frac{k}{K} , \nonumber \\
r_d (k)&=&\frac{n_1-n_{eff}(k)}{2 n_1 \left[1+\alpha(k)\right]}-
i \beta_0  \frac{  n_{eff}(k) }{2  \left[1+\alpha(k)\right]^2} \frac{k}{K}, \quad \quad
\end{eqnarray}
where $n_1=\sqrt{\varepsilon_1}$ and $n_{eff}(k)=\sqrt{\varepsilon_{eff} \mu_{eff}(k)}$.
In the weak-dispersion regime ($k/K \ll 1$), the  magnetic effect is negligible [i.e., $\omega(k)\simeq c k/\sqrt{\varepsilon_{eff}}$] and the temporal transmission and reflection coefficients can be approximated as first-order Maclaurin series in $k$, viz., 
\begin{eqnarray}
\label{tr_quasi}
t_d (k) & \simeq &\frac{1}{2} \left( 1+\frac{\sqrt{\varepsilon_{eff}}}{\sqrt{\epsilon_1}}+ i \beta_0 \sqrt{\varepsilon_{eff}}  \frac{k}{K}      \right), \nonumber \\ 
r_d (k)& \simeq &\frac{1}{2} \left( 1-\frac{\sqrt{\varepsilon_{eff}}}{\sqrt{\epsilon_1}}-i  \beta_0 \sqrt{\varepsilon_{eff}}   \frac{k}{K}      \right).
\end{eqnarray}
From Equations (\ref{tr_quasi}), by taking into account  Equations (\ref{ddd}) and (\ref{cost2}), and maintaining terms up to the first order in $k/K$, we obtain the corresponding expressions for the electric field  
\begin{eqnarray}
\label{tr_ee}
t_e(k) & \simeq &\frac{1}{2} \left( \frac{\varepsilon_1}{\varepsilon_{eff}}+\frac{\sqrt{\varepsilon_{1}}}{\sqrt{\epsilon_{eff}}}+ i \beta_0 \frac{\varepsilon_1} {\sqrt{\varepsilon_{eff}}}   \frac{k}{K} \right), \nonumber \\ 
r_e(k) & \simeq &\frac{1}{2} \left( \frac{\varepsilon_1}{\varepsilon_{eff}}-\frac{\sqrt{\varepsilon_{1}}}{\sqrt{\epsilon_{eff}}}- i \beta_0 \frac{\varepsilon_1} {\sqrt{\varepsilon_{eff}}}   \frac{k}{K} \right),
\end{eqnarray}
where the $k$-dependent terms account for the (weak) dispersion. By neglecting these nonlocal terms,
we recover the well-known  analytical expressions derived in previous studies on conventional (local) temporal boundaries \cite{Xiao}.

\begin{figure*}
\centering
\includegraphics[width=1\textwidth]{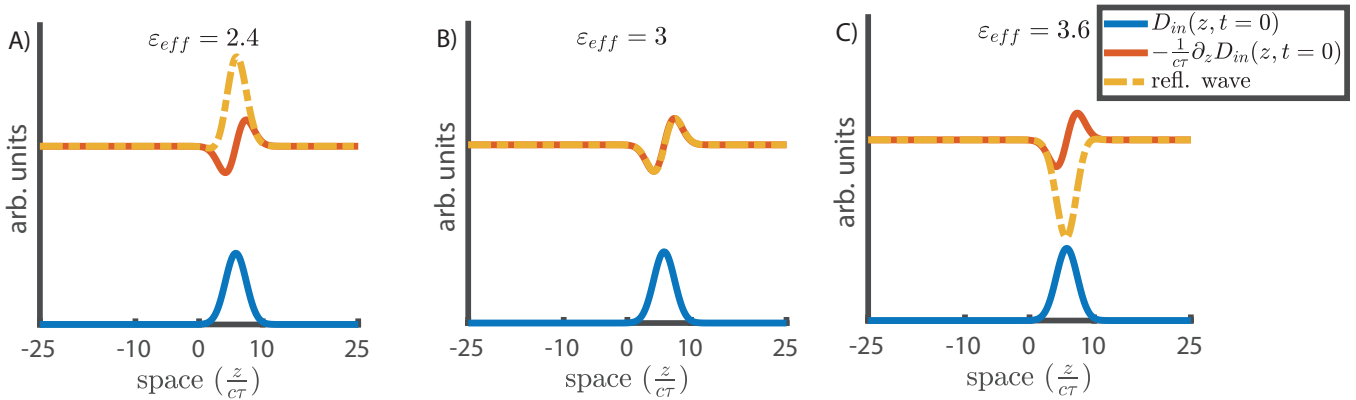}
\caption{Spatial profiles of reflected (backward) pulses $D_r(z,t_1)$ ($t_1 \simeq 20 \tau$) in the configuration described by Equations (\ref{temp}) and (\ref{e(t)}), with $\varepsilon_1=3$, $t_0=10 \tau$, $\phi=3\pi/2$, $\Delta=0.8$, and $\overline{\varepsilon}=4,5,6$ shown in panels (A), (B) and (C), respectively. The blue curve represents the incident pulse profile $D_{in}(z,t=0)$, and the backward pulse $D_r(z,t_1)$ (orange-dashed) is superposed to the first spatial derivative of $D_{in}(z,t=0)$ (red).}
\label{fig_der1}
\end{figure*}
\subsection{Harnessing the temporal-boundary nonlocality}
As for the spatial case \cite{Castaldi}, it is insightful to explore to what extent nonlocality in temporal metamaterials can be harnessed to attain some elementary pulse-shaping or analog-computing functionalities.  
By inspecting Equations (\ref{tr_quasi}) and (\ref{tr_ee}), it is apparent that, in the weak-dispersion regime, backward and forward fields at a nonlocal temporal boundary are dominated by a linear combination of a local (constant) term and a nonlocal correction that is  proportional to $k$, and hence corresponds to a first derivative. Remarkably, in the temporal reflection coefficients $r_d(k)$ and $r_e(k)$, this nonlocal correction can be made dominant by enforcing the impedance-matching condition $\varepsilon_{eff}=\varepsilon_{1}$, so that the local terms vanish. Thus, the backward waveform is proportional to the first derivative of the incident one.

Via Fourier transform, a generic wavepacket can be synthesized as the superposition of the time-harmonic modes in Equation (\ref{epp}), i.e., $D(z,t)=\int_{-\infty}^{+\infty} d(k,t)  e^{i k z}dk$. 
Then,  by recalling Equations (\ref{tr}), we can calculate the forward and backward wavepacket signals. 
\begin{figure*}
\centering
\includegraphics[width=\textwidth]{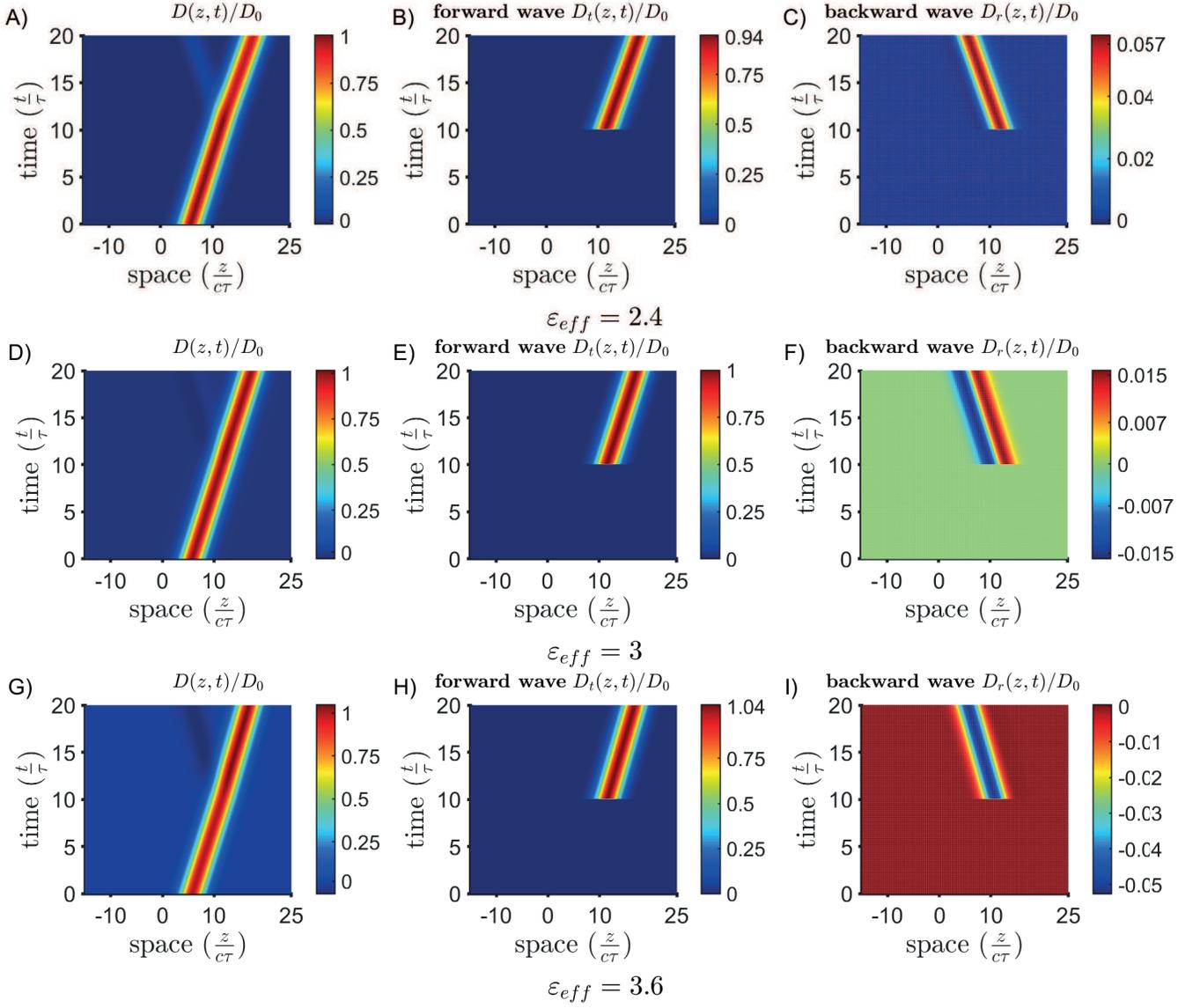}
\caption{As in Figure \ref{fig_der1}, but space-time maps of the electric induction $D$ and the corresponding forward ($D_t$) and backward ($D_r$) components for $\varepsilon_{eff}=2.4$ (A,B,C), 
$\varepsilon_{eff}=3$ (D,E,F), and $\varepsilon_{eff}=3.6$ (G,H,I).}
\label{fig_der2}
\end{figure*}
As a representative example, we consider a Gaussian pulse interacting with the time-varying medium described by Equation (\ref{temp}), where $\varepsilon$ is given by Equation (\ref{e(t)}), with $\phi=3\pi/2$. For $t<t_0$, we assume that the incident pulse propagates in a dielectric medium ($\varepsilon_1=3$) with the following profile 
\begin{equation}
\label{gauss}
{\bf D}_{in}(z,t)=D_0 e^{-\left[\frac{z-v_1 \left(t + t_0 \right)}{v_1 \sigma_t} \right]^2} \hat{\bf e}_x,
\end{equation}
with $v_1= c/ \sqrt{\varepsilon_1}$, $\sigma_t=4 \tau$, $t_0=10 \tau$, and $D_0$ being a constant. After the time instant $t=t_0$, the pulse is partially reflected by the temporal boundary, as illustrated in Figure \ref{fig_der1} (where the dielectric modulation depth is $\Delta=0.8$). It is evident that, for impedance-mismatched configurations, the backward pulses exhibit the same shape as the incident one [see panels (A) and (C), where $\overline{\varepsilon}=4,6$, corresponding to $\varepsilon_{eff}=2.4,3.6$, respectively]. On the contrary, when the impedance is matched [see panel (B), where $\overline{\varepsilon}=5$, corresponding to $\varepsilon_{eff}=3$], the backward pulse reproduces the spatial first derivative of the incident one, as predicted by the second of Equations (\ref{tr_quasi}). For completeness, Figure \ref{fig_der2} shows the space-time maps for the three cases considered:  $\varepsilon_{eff}=2.4$ [panels (A,B,C)], $\varepsilon_{eff}=3$ [panels (D,E,F)], and $\varepsilon_{eff}=3.6$ [panels (G,H,I)]. 

Finally, we validate these predictions from the nonlocal EMT model against full-wave numerical simulations (see the Methods Section \ref{fullwave} for details). Here, we consider the same EM configuration analyzed above with 
$\overline{\varepsilon}=5$, and $\Delta=0.8$, so that the impedance matching condition $\varepsilon_{eff}=\varepsilon_1=3$ is satisfied.

\begin{figure*}
\centering
\includegraphics[width=\textwidth]{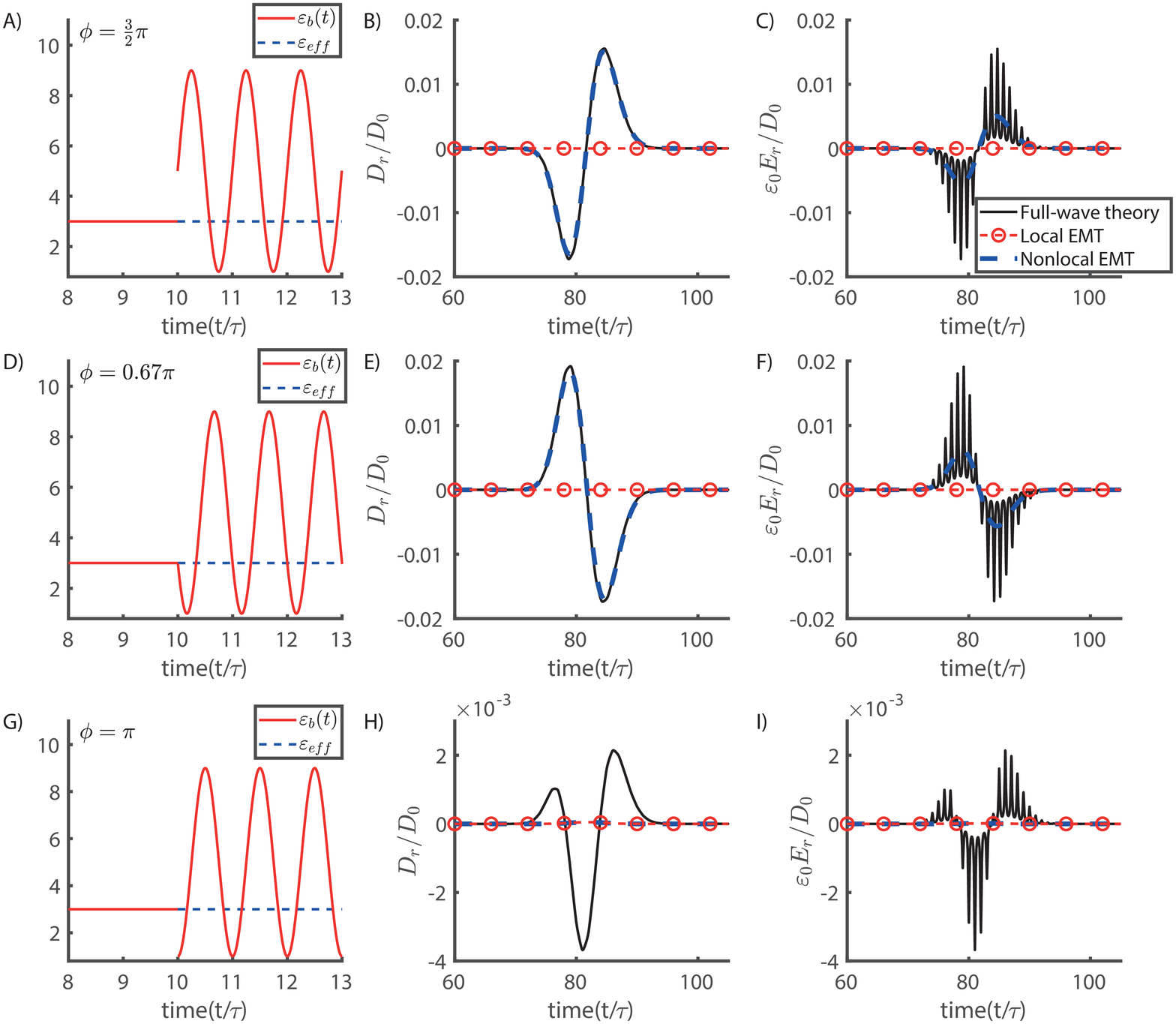}\caption{Comparison between the backward wavepacket profiles predicted by full-wave simulations, local and non-local EMTs. The considered temporal dielectric profiles ensuring the impedance-matching conditions [i.e., the configuration described by Equations (\ref{temp}) and (\ref{e(t)}), with $\varepsilon_1=3$, $t_0=10 \tau$, $\Delta=0.8$, and $\overline{\varepsilon}=5$] is plotted with $\phi=3\pi/2$ (panel A), $\phi=0.67 \pi$ (panel D), and $\phi=\pi$ (panel G). Panels (B),(E), and (H) show the corresponding normalized electric inductions $D_r/D_0$ for the backward pulses. Panels (C), (F), and (I) show the corresponding normalized electric fields $\varepsilon_0 E_r/D_0$. All temporal profiles are evaluated at $z=-30 c \tau$.}
\label{fig_der3}
\end{figure*}

Figure \ref{fig_der3} shows the results on three cases where $\phi=3\pi/2$ (A,B,C), $\phi=0.67 \pi$ (D,E,F), and $\phi=\pi$ (G,H,I). Specifically, we compare the electric induction and electric field distributions predicted by the conventional (local), nonlocal EMT, and full-wave simulations for both cases. 
Panels (B),(E), and (H) show the backward pulse profiles $D_r/D_0$ at $z=-30 c \tau$ for the configurations in panels (A), (D) and (G), respectively. At the impedance matching condition, the local EMT predicts zero temporal reflection, and it does not properly describe the backward wave dynamics. For $\phi=3\pi/2$ and $\phi=0.67 \pi$, our proposed nonlocal EMT is in very good agreement with the full-wave simulations. For  $\phi= \pi$, the local and nonlocal EMT predict zero temporal reflection, whereas full-wave simulations yield a very small backward reflection signal (about an order of magnitude weaker than the previous cases). In this latter case,
the parameter $\beta_0$ ruling the nonlocal effect  vanishes, and the temporal reflection is negligible [see Fig. 3 along with the second of Equations (\ref{tr_quasi})]. Also shown in panels (C), (F) and (I) are the corresponding profiles of the normalized electric fields $\varepsilon_0 E_r/D_0$. In panels (C), (F) and (I), we observe that the full-wave predictions exhibit fast modulations due to the temporal modulation of the permittivity, whereas the nonlocal EMT prediction obtained from Equations (\ref{tr_ee}) is only representative of the slow component. 
Clearly, Equations (\ref{tr_ee}) are the temporal transmission and reflection coefficients of the average component of the electric field, and all fast scales are not considered.

Overall, the above results confirm the validity of the nonlocal EMT predictions to harness the temporal-boundary nonlocality. In the configurations with $\phi=3\pi/2$ and $\phi=0.67 \pi$, the nonlocal contributions, appearing in Equations (\ref{tr_quasi}) and (\ref{tr_ee}), are significant ($\beta_0=0.31,-0.35$ for $\phi=3\pi/2, 0.67 \pi$, respectively), so that these results also confirm the possibility to attain the spatial first derivative of the incident wavepacket. Remarkably, by tailoring the modulation phase $\phi$, the amplitude and the symmetry of the backward pulse can be suitably tuned. Comparing the field profiles in panels (B,C) with those in panels (C,F), it is evident that  the pulse reverses its temporal symmetry profile by switching $\phi$ from $3\pi/2$ to $0.67 \pi$. Quite interestingly, for $\phi=0.67 \pi$, the relative-permittivity function $\varepsilon_b(t)$ is {\em continuous}, thereby indicating that the predicted nonlocal effects do not depend critically on abrupt temporal changes.  

In principle, by exploiting more complex modulation schemes with additional degrees of freedom (e.g.,  temporal dielectric structures with multiple harmonics), it could be possible to tailor the parameters so as to perform higher-derivative orders (and their linear combinations).

%
%
%
%
%
%
%


%

%
%
%
\section{Conclusions}
In summary, via a rigorous multiscale approach, we have developed a nonlocal EMT for temporal metamaterials characterized by permittivity profiles rapidly modulated in time. In analogy with the spatial case, we have elucidated the nonlocal effects, occurring in specific parameter regimes, manifested as an effective diamagnetic response and the possibility to perform basic signal-processing (e.g., first derivative), respectively. In good agreement with full-wave numerical simulations, these results bring about new perspectives and degrees of freedom in the design of temporal metamaterials for tunable nanophotonics and optical computing.

Current and future studies are aimed at exploring more general spatio-temporal modulation schemes, such as  multifrequency and traveling-wave \cite{Huidobro,Galiffi2}. Also crucial from the application viewpoint is the exploration of possible  implementations, based on technological platforms that have been demonstrated at microwave \cite{Kord}, terahertz \cite{Kamaraju}, and optical \cite{Preble} frequencies. Finally, of great interest is a study of the possible effects of topological properties, as in photonic time crystals \cite{Lustig,Ma}, which may enable novel advanced functionalities.

\section{Methods}
\subsection{Nonlocal magnetism}
\label{TOM}
Recalling that the effective Maxwell’s equations are invariant with respect to the Serdyukov-Fedorov transformation \cite{Fed}
${\bf D '}=\overline{\bf D}- \nabla \times {\bf Q}$ and 
${\bf H '}=\overline{\bf B}/\mu_0- \partial_t {\bf Q}$, after setting ${\bf Q}= \partial_t {\bf M}$, we obtain the equivalent effective constitutive relationships
\begin{subequations}
 \begin{eqnarray}
\label{eff4}
{\bf D}' &=& \varepsilon_0 \varepsilon_{eff} {\bf E}' \\
{\bf B}' &=& \mu_0 \left({\bf H'}+ {\bf M} \right), 
\label{eq:Bp}
\end{eqnarray}
\end{subequations}
where 
\begin{eqnarray}
\label{MM}
\nabla^2 {\bf M}-\frac{K^2}{\gamma \varepsilon_{eff}} {\bf M}+\varepsilon_0 \varepsilon_{eff} \frac{\partial^2 {\bf B}'}{\partial t^2} =0,
\end{eqnarray}
 ${\bf E}'=\overline{\bf E}$, and ${\bf B}'=\overline{\bf B}$. In Equation (\ref{eq:Bp}), the vector $\bf M$ plays the role of an effective magnetic polarization. Therefore, in an isotropic and spatially homogeneous temporal metamaterial, the spatial dispersion up to the second-order (i.e., up to $\eta^2$) is {\em fully equivalent} to optical magnetism with the magnetic polarization given by Equation (\ref{MM}). Considering the propagation of a monochromatic plane wave, we obtain that the EM fields experience a nonlocal magnetic response described by the effective relative magnetic permeability given in Equation (\ref{muu}).
\subsection{Rigorous dispersion relationship in time-periodic varying media}
Following the rigorous approach in \cite{Zurita}, we focus on the wave equation describing the electric field dynamics in a time-periodic varying medium. The propagation of a plane wave ${\bf E}(z,t)=2 {\rm Re} \left[ e(t) e^{ikz}  \right] \hat{\bf e}_x$ is described by the equation 
\begin{equation}
\label{eq_exa}
\frac{d^2 }{d t^2}[\varepsilon(t) e(t)]+k^2 c^2 e(t)=0, 
\end{equation}
where $\varepsilon(t)=\varepsilon(t+\tau)$.  
Since the permittivity is periodic in time, this  equation admits Bloch-type modes $e(t)={\tilde e} (\omega,t) e^{-i \omega t}$, where  ${\tilde e} (\omega,t)$ is a periodic function of period $\tau$. By expanding in Fourier series the observables, Equation (\ref{eq_exa}) becomes 
\begin{equation}
\label{eq_exb}
\sum_{n} \left[ (\omega-\Omega m)^2 {\tilde c}_{m-n}-k^2 c^2\delta_{m,n} \right] {\tilde e}_n=0, 
\end{equation}
where ${\tilde c}_{n}$, ${\tilde e}_n$ are the Fourier coefficients for the relative permittivity and electric field, respectively, and $\delta_{m,n}$ is the standard Kronecker-delta tensor ($m,n=0,\pm 1,\pm 2,...$).  
Equation (\ref{eq_exb}) is a set of linear equations that exhibits a nontrivial solution only if the associated determinant vanishes. By suitably truncating the Fourier series expansions, we numerically obtain the rigorous dispersion relationship $\omega(k)$ for waves propagating in a periodic temporal medium. 
\subsection{Full-wave simulations}
\label{fullwave}
 We consider an arbitrary wavepacket propagating in a spatially homogeneous unbounded, time-varying metamaterial, with the temporal boundary described by Equation (\ref{temp}).
The wavepacket electric induction can be synthesized via Fourier transform as
\begin{equation}
\label{TF}
{\bf D} \left(z,t\right)= \hat{\bf e}_x \int_{-\infty}^{\infty}d\left( k,t \right) {e}^{i k z} dk.
\end{equation}
Then, for each value of the wavenumber $k$, we define the auxiliary functions
\begin{equation}
u_1\left(k,t\right)=\frac{d \left(k,t\right)}{D_0},\quad \ u_2\left(k,t\right)=\frac{b \left(k,t\right)}{D_0 Z_0},
\end{equation}
where $d\left(k,t\right)$ and $b\left(k,t\right)$ are the plane-wave spectra of the electric and magnetic inductions, respectively,  at a given time $t$, with $Z_0$ and $D_0$ denoting the vacuum intrinsic impedance and a dimensional normalization constant, respectively. From Maxwell's curl equations, we derive a pair of coupled ordinary differential equations, namely
\begin{eqnarray}
\frac{du_1 }{dt}&=&-ick u_2,\nonumber\\
\frac{du_2}{dt}&=&-ick\frac{u_1}{\varepsilon_b},
\label{system}
\end{eqnarray}
with initial conditions
\begin{equation}
	u_1\left(k,0\right)=\frac{d_{in}(k,0)}{D_0}, \quad \quad u_2\left(k,0\right)=\frac{d_{in}(k,0)}{D_0 \sqrt{\varepsilon_1}},
\end{equation}
where $d_{in}(k,0)$ is the plane wave spectrum of the incident electric induction field at $t=0$.
Next, we solve numerically Equations (\ref{system}) by means of the \texttt{NDSolve} routine available in Mathematica\texttrademark \cite{Mathematica}. This routine provides the numerical solution of generic systems of ordinary differential equations, via a broad arsenal of methods (including Runge-Kutta, predictor-corrector, implicit backward differentiation) that can be tailored adaptively to the specific scenario of interest and, in principle, it can automatically handle discontinuities in the equations \cite{Mathematica}.
In our implementation, we utilize default settings and parameters. Moreover, in order to favor numerical convergence, we implement the abrupt permittivity changes via an analytical, smooth unit-step function $U_s\left(t\right) = \left[\tanh\left(t/T_s\right) + 1\right]/2$, where $T_s=\tau/100$.
 
Once a numerical solution is available for Equations (\ref{system}), the electric induction is synthesized via Equation (\ref{TF}) (where $d=D_0 u_1$), whereas the corresponding electric field can be readily obtained via division by $\varepsilon_b\left(t\right)$. In our numerical implementation, this synthesis is implemented via fast-Fourier-transform by means of the  \texttt{Fourier} routine available in Mathematica\texttrademark \cite{Mathematica}.

\section*{ACKNOWLEDGEMENT}

G.C. and  V.G. acknowledge partial support by the University of Sannio via the FRA 2020 program. 


\end{document}